\documentstyle[aasms4]{article}

\def\ltsima{$\; \buildrel < \over \sim \;$}
\def\simlt{\lower.5ex\hbox{\ltsima}}
\def\gtsima{$\; \buildrel > \over \sim \;$}
\def\simgt{\lower.5ex\hbox{\gtsima}}

\def\ls{{_<\atop^{\sim}}}
\def\gs{{_>\atop^{\sim}}}

\begin{document}

\title{Broad Emission Line Regions in AGN: the Link 
with the Accretion Power} 

\author{Fabrizio Nicastro$^{1,2,3}$}

\affil {$^1$ Harvard-Smithsonian Center for Astrophysics, 
60 Garden st. Cambridge Ma. 02138 USA}
\affil {$^2$ Osservatorio Astronomico di Roma, 
via Osservatorio, Monteporzio-Catone (RM), I00040 Italy}
\affil {$^3$ Istituto di Astrofisica Spaziale - CNR, 
Via del Fosso del Cavaliere, Roma, I-00133 Italy}

\today

\begin{abstract}
We present a model which relates the width of the Broad Emission 
Lines of AGN to the Keplerian velocity of an accretion disk at 
a critical distance from the central black hole. 
This critical distance falls in a region bounded on the inward side 
by the transition radius between the radiation pressure and the gas 
pressure dominated region of the accretion disk and on the outward side 
by the maximum radius below which a stabilizing, radially accreting and 
vertically outflowing corona exists.\\
We show that in the framework of this picture the observed range of 
H$\beta$ FWHM from Broad Line to Narrow Line type 1 AGN is well 
reproduced as a function of the accretion rate. This interval of 
velocities is the only permitted range and goes from $\sim 20,000$ 
km s$^{-1}$ for sub-Eddington accretion rates, to $\sim 1,000$ km 
s$^{-1}$ for Eddington accretion rates.  
\end{abstract}


\section{Introduction}
Broad Emission Lines (BELs) are probably ubiquitous in AGN, 
being unobserved only when a good case can be made for their 
obscuration by dust (type 2 AGN) or their being swamped by beamed 
continuum (Blazars). This suggests that the presence of Broad 
Emission Line Clouds (BELCs) in the AGN environment is closely 
related to the mechanisms which are responsible for the quasar 
activity. 
Shakura \& Sunyaev (1973, hereinafter SS73) already noted that a 
high velocity, high density wind of ionized matter may originate 
from the accretion disk (a suggestion greatly elaborated by Murray 
et al., 1995), at a radial distance where the radiation pressure 
becomes comparable to the gas pressure. 
Witt, Czerny and Zycki (1997, hereinafter WCZ97) studied a radially 
co-accreting disk/corona system (which stabilizes the, otherwise 
thermally unstable, radiation pressure dominated region of a 
Shakura-Sunyaev disk - SS-disk), and demonstrated that a 
transonic vertical outflow from the disk develops where 
the fraction of total energy dissipated in the corona is close 
to the maximum. 
In a revised version of this model (Czerny et al., 1999, in
preparation), which includes evaporation from the disk, this solution 
continues to hold for accretion rates higher than a minimum value, 
below which evaporation inhibits the formation of the wind. 

Reverberation studies of the BELs in many Seyfert 1 galaxies 
(e.g. Wandel, Peterson \& Malkan, 1999: hereinafter WPM99) 
indicate that the gas of the BELCs is photoionized, and that 
its physical state, from one object to another, covers a rather 
narrow range of parameter space (in column density, electron 
density, ionization parameter). It is also clear (Krolik et al., 
1991; Peterson et al., 1999) that within a single object the 
BELCs are stratified, with the highest ionization lines being 
also the broadest. This accords with the photoionization hypothesis 
and with the idea that the BELs are broadened by their orbital 
Keplerian motion around the central source (Peterson \& Wandel, 1999). 

On the other hand the BELs do not at all have the same dynamical 
properties in all objects. A broad distribution of line widths 
from $\sim 1,000$ km s$^{-1}$ (in Narrow Line Seyfert 1 galaxies, 
NLSy1) to $\sim 20,000$ km s$^{-1}$ (in the broadest broad line type 
1 AGN) is present. 
A model by Wandel (1997) attempts to forge a physical link between 
the breadth of the permitted optical emission lines of type 1 AGN 
and the steepness of their X-ray continuum: a steeper 
X-ray spectrum has stronger ionizing power, and hence the BELR 
is formed at a larger distance from the central source, where the velocity 
dispersion is smaller, and so produces narrower emission lines. 
Alternatively, Laor et al. (1997) use a dynamical argument to suggest 
that small emission line widths are direct consequence of large 
L/L$_{Edd}$. 

\medskip
Here we propose that a vertical disk wind, originating at a critical 
distance in the accretion disk, is the origin of the BELCs and 
that the widths of the BELs are the Keplerian velocities of the 
accretion disk at the radius where this wind arises. The disk wind 
forms for external accretion rates higher than a minimum value below 
which a standard SS-disk (SS73) is stable and extends down to the last 
stable orbit. 
The model explains the observed range of FWHM 
in the BELs of AGN as a function of a single physical quantity 
connected with the AGN activity: the accretion rate. 
In \S 2 we present our model, and show the basic equations 
which support our findings. In \S 3 we discuss the observational 
consequences and compare with exisiting data. 

\section{The Model}
The three main ingredients of our model are: (a) the transition 
radius $r_{tran}$, derived by setting equal the radiation pressure 
at $r < r_{tran}$ and the gas pressure at $r > r_{tran}$, in a standard 
SS-disk (SS73): 
\begin{equation} \label{rtran}
r_{tran} f^{-16/21} \simeq 15.2 (\alpha m)^{2/21} \left({1 \over \eta} 
\dot{m} \right)^{16/21}, 
\end{equation}
(b) the approximate analytical relationship giving the fraction of 
energy dissipated in the corona, in a dynamical disk/corona configuration 
(WCZ97): 
\begin{equation} \label{beta}
(1-\beta) \simeq 0.034 (\alpha f {1 \over \eta} \dot{m})^{-1/4} r^{3/8}, 
\end{equation}
and (c) the maximum radius below which a stable co-accreting disk/corona 
configuration can exist, obtained by setting $\beta = 0$ (WCZ97): 
\begin{equation} \label{rmax}
r_{max} f^{-2/3} \simeq 8,000 \left( \alpha {1 \over \eta} 
\dot{m} \right)^{2/3}. 
\end{equation}
In the above equations we have used dimensionless quantities: $m = 
M/M_{\odot}$, $\dot{m} = \dot{M} / \dot{M}_{EDD}$, $r = R/R_0$, 
with $\dot{M}_{EDD} = 1.5 \times 10^{17} \eta^{-1} m \hbox{ }$ g 
s$^{-1}$, and $R_0 = 6GM/c^2$, for a non-rotating black hole; here 
$\eta$ is the maximum efficiency. Finally, $f$ gives the boundary 
conditions at the marginally stable orbit: $f = f(r) = (1 - r^{-0.5})$. 

We note that (with the adopted units) $r_{max}$ does not depend on the 
mass, while $r_{tran}$ depends only very weakly on it. 
However both these critical radii depend on the accretion rate and, 
interestingly, with similar powers. This results in a quasi-rigid 
radial shifting of the region delimited by these two distances 
as the accretion rate (in critical units) varies. 
From equations \ref{rtran} and \ref{rmax} we can estimate  
the total radial extent of this region to be of the order of 
$\sim 10$ times $r_{tran}$, for $m = 10^8$ and $\dot{m} = 1$ 
\footnote{
The radial extent of the vertical outflow containing the BELCs, however, 
is smaller than this, being constrained by the weighting function 
$(1 - \beta)$ (equation \ref{beta}, WCP97. See below).}. 
Equation \ref{rtran} allows us to define the minimum 
external accretion rate needed for a thermally unstable radiation 
pressure dominated region to exist. 
From the condition $r > 1.36$ (the limit of validity of the SS-disk 
solution) we have: $\dot{m} \gs \dot{m}_{min}(m) 
\simeq 0.3 \eta (\alpha m)^{-1/8}$. 
Throughout this paper we assume $\eta = 0.06$, and a viscosity 
coefficient of $\alpha = 0.1$, which give a minimum external accretion 
rate of $\dot{m}_{min} \sim (1-4) \times 10^{-3}$, for $m$ in 
the range $10^6 - 10^9$. 
At lower accretion rates a SS-disk (SS73) is stable down to the 
last stable orbit: we propose that all the available 
energy is dissipated in the disk and no radiation pressure 
supported and driven wind is generated. AGN accreting at these low 
external rates should show no BELs in their optical spectra. 
For accretion rates $\dot{m} \gs \dot{m}_{min}$ a SS-disk is 
unstable (Lightman \& Eardley, 1974) and a stabilizing, 
co-accreting ``disk/corona + outflow'' system forms (WCZ97). 
The fraction of energy dissipated in the corona, and powering the 
vertical outflow, is maximum at $r_{max}$ and decreases inward following 
equation \ref{beta} (see also Fig. 6 in WCZ97). 
At radii smaller than $r_{tran}$ the available energy is almost 
equally divided between the disk and the corona (WCZ97). 
We then adopt an averaged radius $r_{wind}$ for the transonic 
outflow (and so for the BELRs), obtained weighting the radial distance 
by $(1-\beta)$ - equation \ref{beta} - , between $r_{tran}$ and $r_{max}$. 
We computed $r_{wind}$ numerically for several values of $m$ and 
$\dot{m}$. 

\medskip
In the next section we discuss the implications of our model for 
linking the gas of the BELR in type 1 AGN and the accretion 
mechanism. 

\subsection{Dynamical Properties}
We calculated the orbital velocities at $r_{wind}$ under the Keplerian 
assumption: $\beta(r_{wind}) = v/c = (6 r_{wind})^{-1/2}$. 
We then transformed these velocities to Full Width Half Maximum (FWHM) 
values of the lines emitted by these clouds of gas using the relationship: 
FWHM$ = 2 (<v^2>)^{1/2}$ (Netzer, 1991), where $(<v^2>)^{1/2} = 
v/\sqrt{2}$ is the averaged Keplerian velocity in a cylindrical 
geometry. 
In Figure 1 we show the relationship between the accretion rate (in the 
range $\dot{m}_{min}(m) - 10$) and the expected FWHM$(r_{wind})$ 
(solid, thick curves) and FWHM($r_{max}$) (dashed, thin curves), for 
$m = 10^6, 10^7, 10^8, 10^9$. 
These curves are independent on the mass of the 
central black hole, and so they overlap in the diagram of 
Figure 1. 
However, for each curve, the maximum FWHM reacheable depends on the 
mass, through the limit imposed by the minimum external accretion 
rate $\dot{m}_{min}(m)$ needed for an unstable SS-disk to exist. 
At the bottom of the plot the four horizontal lines indicate these 
values of $\dot{m}_{min}(m)$. 
The parameter space delimited by the dashed and solid curves of Figure 1 
gives a possible range of FWHM at a given accretion rate, so allowing, 
at least partly, for a stratification of the BELCs in a single object
\footnote{However, according to the model for the quasar structure 
proposed by Elvis (1999), a more complete and satisfactory explanation 
is that the high and low ionization BELs are actually produced in two 
separate regions of the outflow: in the vertical part, at a height 
$z < r$ from the disk surface (high ionization lines), and in the 
radially displaced part (low ionization lines) located at $z > r$ 
(WCZ97) and shadowed by the vertically flowing gas (Elvis, 1999).}. 
Finally, we marked two different regions in the diagram of Figure 2: 
(a) for accretion rates $\dot{m} < 0.2$ (sub-Eddington regime) the 
predicted FWHM are quite broad ($\gs 4,000$ km s$^{-1}$), and similar 
to those typically observed in broad line type 1 AGN; (b) for $\dot{m} 
= 0.2 - 3$ (Eddington to moderately super-Eddington) the corresponding 
FWHM span the interval $\sim 1,000 - 4,000$ km s$^{-1}$, which 
contains the value of FWHM$ = 2,000$ km s$^{-1}$ used to separate the two 
classes of broad line type 1 AGN and NLSy1. 
Hence our model predicts that narrow line type 1 AGN accrete at 
higher accretion rates compared to broad line objects, as in the 
Pounds et al. (1995) suggestion. 
However, the mass of the central black hole in NLSy1 does 
not need to be smaller than that of broad line type 1 AGN, which reconciles 
the NLSy1 paradigm with the recent results of WPM99 and the mass 
estimate for the NLSy1 galaxy TON~S180 (Mathur S., private 
communication). 

\section{Comparison with Observations}
Figure 3 shows an analogous and complementary diagram to 
Figure 2, where we plot $r_{wind}$ in physical units on the Y-axis. 
The four mostly horizontal curves correspond to the four black hole 
masses $m = 10^6, 10^7, 10^8, 10^9$, with the accretion rate increasing 
from the bottom-right to the top-left of the diagram. 
The four vertical lines correspond to four values of the accretion 
rate $\dot{m} = 0.01, 0.1, 1, 10$, with black hole mass increasing 
from the bottom to the top. 
The space delimited by this grid contains all the allowed range of 
distances and orbiting velocities of the BELCs in this model. 
We also show a horizontal band which delimits the typical observed 
BELR sizes (WPM99). 
Superimposed on this diagram, we plot the recent measurements of 
distance and FWHM reported by WPM99 and obtained by using the 
reverberation-mapping technique. 
The points are numbered following the order of Table 2 in WPM99. 
For each point the grid of Figure 2 allows one to uniquely determine 
the predicted accretion rate and black hole mass. 
We calculated the ratio $\xi$ between the measured (WPM99) 
dimensionless luminosity $\ell = (L_{ion}/L_{Edd})$ and the predicted 
accretion rate. 
$\xi$ is then a measure of the relative efficiency (compared to the 
maximum radiative efficiency $\eta$: $\xi = \epsilon / \eta$, where 
$\epsilon = L_{ion}/\dot{M}c^2$) with which the accretion power is 
converted into ionizing luminosity.
We found that $\xi$ is correlated with the measured mass of the central 
black hole: $\xi = 10^{-(0.4 \pm 0.9)} \times M_8^{(1.00 \pm 0.14)}$, 
R = 0.68, corresponding to a probability of P($>$R;N=16) = 0.4 \% 
(Figure 3. We do not consider here the tree objects of the WPM99 sample 
for which only upper limits on the central black hole mass were available). 
The correlation is still significant (R = 0.62, P($>$R;N=15) = 1.3 \%), 
when the object (NGC~4051) with the lowest mass and relative radiative 
efficiency is removed. 
This observational result is not an obvious consequence of our model 
and needs study. 
This correlation may explain why we do not see very-low mass 
AGN ($\ls 10^4$ M$_{\odot}$): the efficiency in converting accretion 
power into luminosity may be too low for such objects. 

\section{Testing the Model}

\subsection{Observed Broad Line Widths}
In the framework of our model, those AGN showing particularly broad 
emission lines (FWHM$\sim 15,000-20,000$ km s$^{-1}$) in their optical 
spectra are objects which are accreting at very low rate, close 
to but higher than $\dot{m}_{min}(m)$. 
Their expected ionizing luminosity is then given by 
$L_{ion} \sim \xi(m) \dot{m}_{min}(m) L_{Edd} \sim 10^{43} M_8^{15/8} 
\hbox{ erg s}^{-1}$, 
where $M_8$ is the mass of the central black hole in 
$10^8 M_{\odot}$, and we used the correlation $\xi \sim 0.4 M_8$. 
The predicted ionizing luminosity at the minimum accretion rate are then 
$L_{ion}(M_8=1) \sim 10^{43}$ erg s$^{-1}$ and $L_{ion}(M_8=0.1) 
\sim 10^{41}$ erg s$^{-1}$. 
AGN with strong and exceptionally broad emission lines (e.g. Broad Line 
Radio Galaxies, Osterbrock, Koski \& Phillips, 1975; Grandi \& Pillips, 
1979) should then host a massive central black hole, of $10^8-10^{9}$ 
solar masses. 
Lower mass black holes accreting at rates slightly higher than 
$\dot{m}_{min}(m)$ may also have very broad optical emission lines, but 
they would be hard to detect due to their low contrast optical 
spectra. 

Emission lines broader than $\sim 20,000$ km s$^{-1}$ should 
not exist for any plausible mass of black hole. This limit seems 
to be obeyed. FWHM $\sim 20,000$ km s$^{-1}$ are rare. Steiner (1981) 
lists 14 out of 147 AGN with FW-Zero-Intensity between $20,000$ and 
25,000 km s$^{-1}$ (only 3 with FW-Zero-Intensity $> 23,000$ km s$^{-1}$). 
The number with FW-Half-Maximum $\sim 20,000$ km s$^{-1}$ will be 
much less. 

\subsection{Low Luminosity AGN}
Nearby AGN with independently measured masses (see Ho, 1999) 
that imply accretion at rates lower than $\dot{m}_{min}(m)$ should show 
no broad emission lines in their optical spectra. 

NGC~4594 is a low luminosity ($L_X \sim 3.5 \times 10^{40}$, Nicholson 
et al., 1998) AGN/LINER with a spectroscopically well estimated mass 
of $10^9$ solar masses for the central object (Kormendy et al., 1996). 
This gives a ratio between the X-ray (ASCA) and the Eddington 
luminosity of NGC~4594 of $3 \times 10^{-7}$. In our model no BELR 
should exist at this low $L/L_{Edd}$. 
Nicholson et al. (1998) showed that no broad H$\alpha$ was present 
in the HST spectrum of NGC~4594. 
From the ASCA spectrum they also put an upper limit on the column 
density of cold absorbing gas of $2.9 \times 10^{21}$ cm$^{-2}$. 
This is much lower than the amount of gas which would be needed 
to obscure the putative BELR.
Nicholson et al. (1998) conclude that there is no BELR in NGC~4594. 
All this is in good agreement with the predictions of our model 
(see \S 4): if the object has a mass of $10^9$ M$_{\odot}$, then 
the expected luminosity for the minimum accretion rate $\dot{m}_{min}(m)$ 
is $\sim 7\times 10^{44}$ erg s$^{-1}$. 
An accretion rate of $\dot{m} \sim 5 \times 10^{-5} \times 
\dot{m}_{min}(m)$ would then be necessary to obtain the observed 
X-ray luminosity of $L_X \sim 3.5 \times 10^{40}$ erg s$^{-1}$ 
(Nicholson et al., 1998), at which no BELR is predicted to exist. 

Ho et al. (1997) searched for BELRs in a sample of 486 bright 
northern galaxies. They found that among the 211 objects classified 
as having a Seyfert or LINER nucleus, only 46 ($\sim 20$ \%) have a 
detectable broad emission line component. Among these 46 the maximum 
H$\alpha$ measured is 4,200 km/s. 
The $\sim 80$ \% of the sources in the sample with no broad 
H$\alpha$ are either (a) obscured (type 2) AGN, (b) low massive AGN 
with accretion rates slightly higher than $\dot{m}_{min}(m)$ and in 
which the broad H$\alpha$ component is too broad to be measured in 
low-contrast spectra, or (c) AGN with high mass (and so visible) but 
with an accretion rate lower than $\dot{m}_{min}(m)$, and so with no BELRs 
(like NGC~4594). 
The remaining $\sim 20$ \% of the sample with broad H$\alpha$ would 
be made of very low-mass AGN ($\sim 10^5$ M$_{\odot}$), with 
quite normal (Sy1-like) accretion rate ($\dot{m} \sim 0.1-0.5$), but 
(because of their low mass) with low-luminosity. 
The BELs in these objects would then be quite normal, and more 
easily detectable even in low contrast spectra. 

\section{Conclusions}
We presented a simple model which tightly links the existence of the 
BELCs in type 1 AGN to the accretion mechanism. 
We derived the accretion rate and mass scaling laws for a mean distance 
on a stable SS-disk falling within the region delimited inward by the 
transition radius between the radiation pressure and the gas pressure 
dominated regions, and, outward, by the maximum radius below which a 
stabilizing co-accreting corona forms (WCZ97). We identify the BELR with 
a vertically outflowing wind of ionized matter which forms at this radius. 

Our main findings are: 

\begin{itemize}

\item{} The Keplerian velocity of the BELCs around the 
central black hole depends critically on the accretion rate. 
The entire observed range of velocities (FWHM) 
in type 1 AGN is naturally reproduced in our model 
allowing the accretion rate to vary from its minimum permitted 
value to super-Eddington rates. 


\item{} For accretion rates close to the Eddington value, the 
expected FWHM are of the order of those observed in NLSy1. 
Lower accretion rates give instead FWHM typical of broad line 
type 1 AGN. For very low accretion rates the BELCs would not 
longer exist giving an upper limit to the allowed FWHM of the 
BELs, consistent with observations. This limit could explain the 
absence of BELs in Low-Luminosity AGN. Existing optical data of 
LINERs are consistent with this prediction.   

\item{} In physical units, the distance at which the BELCs 
would form, is a steep function of both the accretion rate 
and the mass of the central object. The predicted relationsips 
agree with the observed masses, FWHM, and distances (WPM99). 

\item{} We find an empirical relationship which suggests that 
the radiative efficiency of higher mass black holes is greater than 
that for lower mass black holes. 

\end{itemize}

\begin{acknowledgements} 
This work has been made possible through the very useful and 
fruitful discussion with my colleagues at SAO and in Rome. 
In particular I would like to thank M. Elvis, F. Fiore and A. 
Siemiginowska for the very stimulating inputs that they gave to 
the development of the ideas presented in this paper. 
A particular thank goes to K. Forster, A. Fruscione, S. Mathur 
and B. Wilkes (at SAO), and G. Matt and G.C. Perola (at the 3rd 
Univesrsity of Rome). 
This work has been partly supported by the NASA grants ADP NAG-5-4808, 
NAG-5-3039, NAG-5-2476 and LTSA NAGW-2201. 
\end{acknowledgements}





\newpage
%
\figcaption{Predicted accretion rate vs FWHM$(r_{wind})$, FWHM($r_{max}$), 
for $m = 10^6, 10^7, 10^8, 10^9$, and $\dot{m}$ in the range 
$\dot{m}_{min}(m) - 10$.}
%

%
\figcaption{$r_{wind}$ vs FWHM($r_{wind}$) curves, for 4 values 
of the black hole mass ($m = 10^6, 10^7, 10^8, 10^9$: mostly horizontal 
curves), and 4 values of the accretion rate ($\dot{m} = 0.01, 0.1, 1, 
10$: vertical curves). 
The units on the Y-axis are both in light-days (left axis) and cm 
(right axis). The points are the measured BELR distances and H$\beta$ 
FWHM for the 19 type 1 AGN of the WPM99 sample.}
%

%
\figcaption{Correlation between the relative radiative efficiency 
$\xi$ and the measured black hole mass $m_{meas.}$.}
%

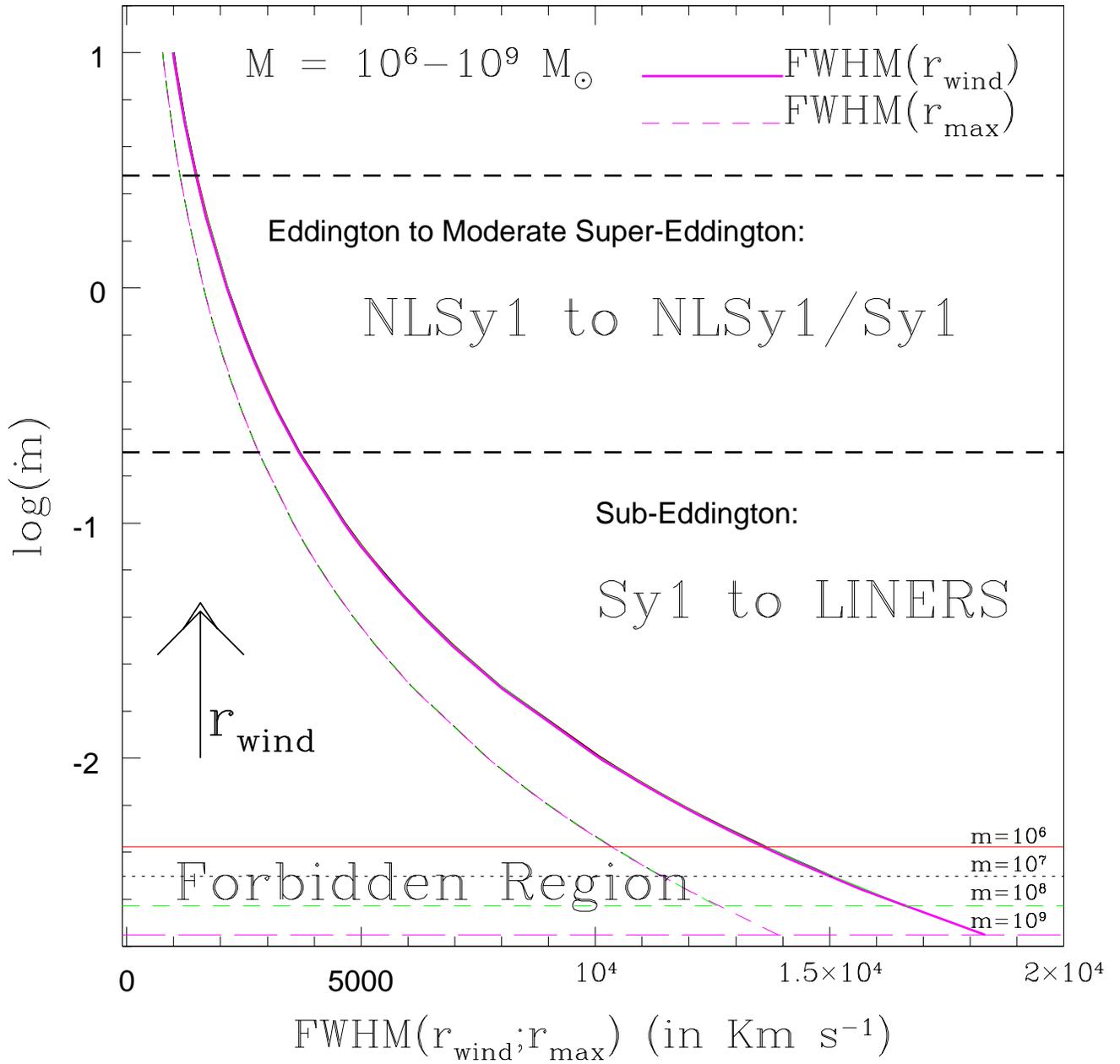
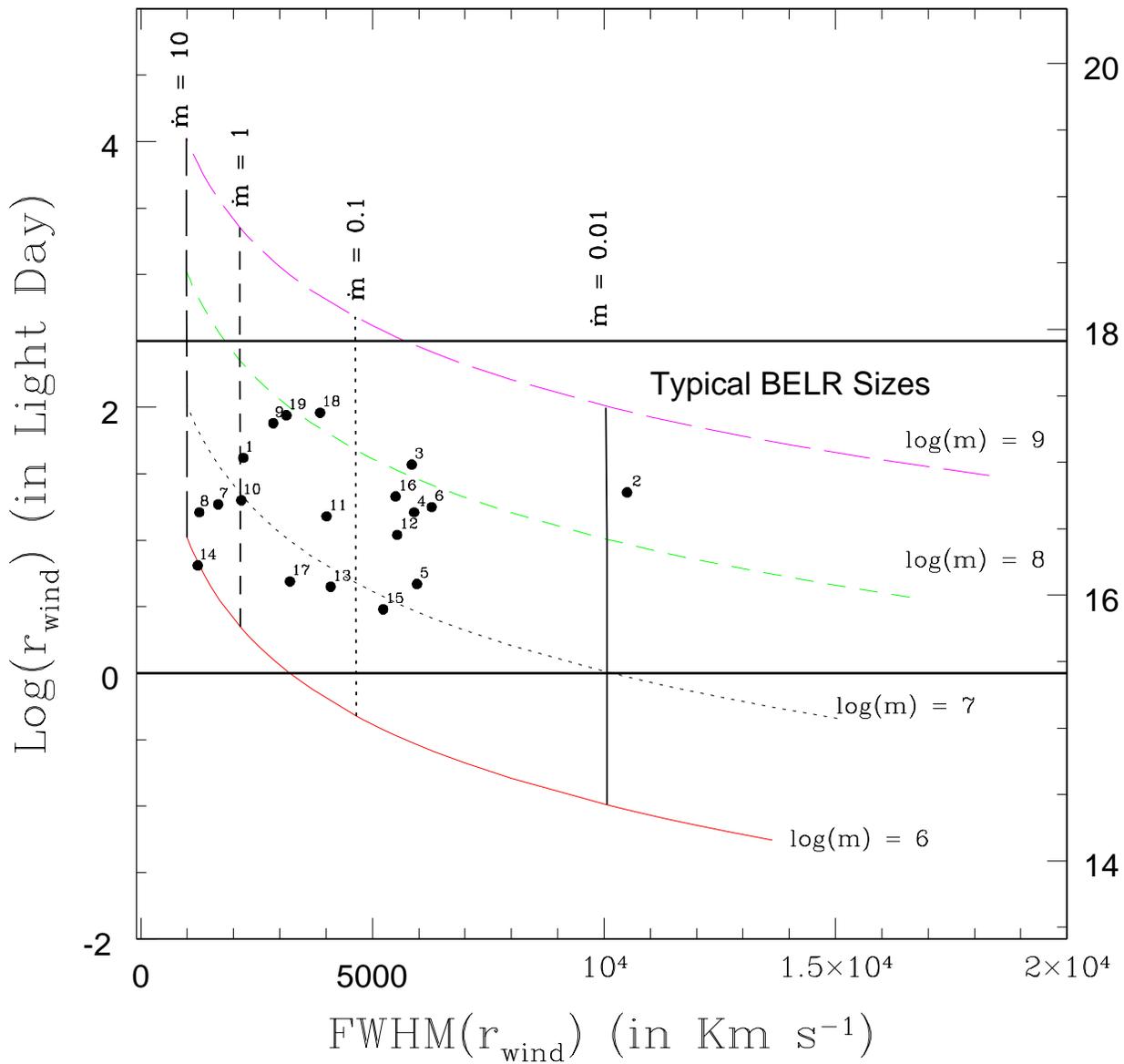
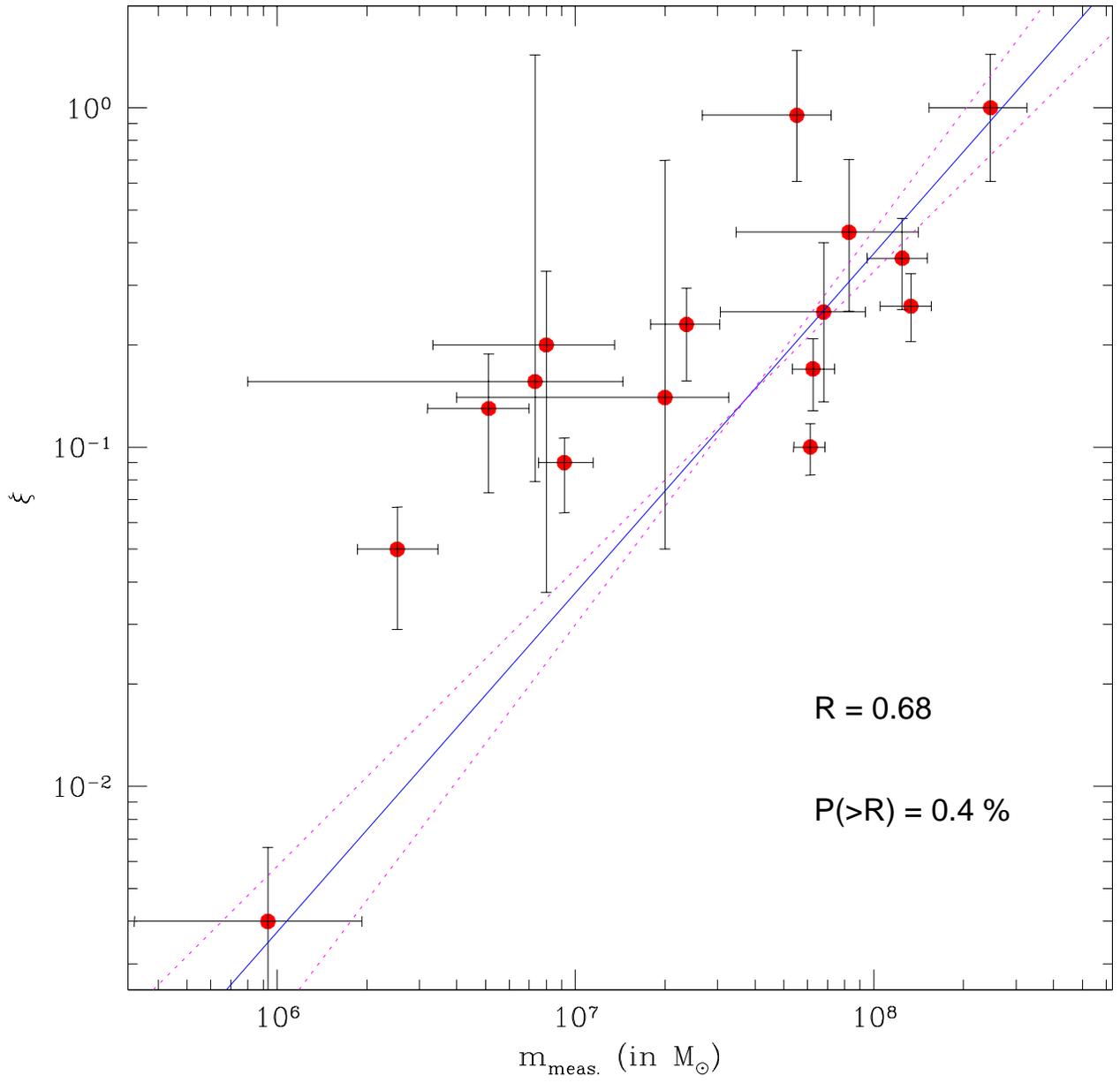
\end{document}